\documentclass[aps,prb,reprint,superscriptaddress,nofootinbib]{revtex4-2}
\usepackage[utf8]{inputenc}

\usepackage[hidelinks]{hyperref}

\usepackage{hyperref}
\usepackage{color}
\usepackage[caption=false]{subfig}

\hypersetup{colorlinks=true}
\usepackage{graphicx}
\usepackage{bm}
\usepackage{amsmath}
\usepackage{amssymb}
\usepackage{xspace}
\usepackage[capitalise]{cleveref}
\usepackage{txfonts}
\usepackage{physics}
\usepackage{siunitx}
\usepackage{tikz}
\usetikzlibrary{external}
\tikzset{external/force remake}
\usepackage{chemformula}

\usepackage[textsize=footnotesize]{todonotes}
\newcommand{\Ih}[1][]{I^{#1}_p(\theta,\dot{\phi})}
\newcommand{\Ib}[1][]{I_V^{#1}(\theta)}
\newcommand{\ii}{\mathrm{i}}

\newcommand{\Vcone}{V_{c,1}}
\newcommand{\Vctwo}{V_{c,2}}

\newcommand{\Hdot}{H_{{n}}}
\newcommand{\bath}{{f}}
\graphicspath{{./figures/}}
\newcommand{\subfigref}[2]{Fig.~\hyperref[#1]{\ref*{#1}#2}}

\newcommand{\rf}{\mathrm{rf}}

\begin{document}
	
	\title{Nonequilibrium effects in spin-torque oscillators}

	\author{Pieter M. Gunnink}
	\email{pgunnink@uni-mainz.de}
	\affiliation{Institute of Physics, Johannes Gutenberg-University Mainz, Staudingerweg 7, Mainz 55128, Germany}

	\author{Tim Ludwig}
	\affiliation{Department of Philosophy, Institute of Technology Futures, Karlsruhe Institute of Technology, Douglasstraße 24, 76133 Karlsruhe, Germany}
	\author{Alexander Shnirman}
	\affiliation{Institute for Theory of Condensed Matter, Karlsruhe Institute of Technology (KIT), 76131 Karlsruhe, Germany}
	\affiliation{Institute for Quantum Materials and Technologies, Karlsruhe Institute of Technology (KIT), 76131 Karlsruhe, Germany}
	\author{Rembert A. Duine}
	\affiliation{Institute for Theoretical Physics and Center for Extreme Matter and Emergent Phenomena, Utrecht University, Leuvenlaan 4, 3584 CE Utrecht, The Netherlands}
	\affiliation{Department of Applied Physics, Eindhoven University of Technology, P.O. Box 513, 5600 MB Eindhoven, The Netherlands}
	\date{\today}
	\begin{abstract}
		One of the cornerstones of spintronics is the application of a spin-transfer torque to a nanomagnet, driving the magnetization of the nanomagnet into a steady-state precession and realizing a spin-torque oscillator.
		Such a steady state, sustained by a balance between driving and dissipation, could be a textbook example for a nonequilibrium situation. Nevertheless, most theoretical descriptions of spin-torque oscillators simply assume local equilibrium. Here, based on a simple model, we investigate the relevance of nonequilibrium effects in spin-torque oscillators. 
		We use a nonequilibrium Keldysh description, which allows us to treat the effects of spin relaxation, and find that, in the absence of spin relaxation, persistent precessions of the magnetization are not allowed, if magnetic anisotropies are absent. 
		However, introducing spin relaxation enables persistent precessions, where the strength of the spin relaxation has quantitative and qualitative effects on the magnetization dynamics.
		In the presence of magnetic anisotropy, we find that persistent precessions are allowed even if spin relaxation is absent, but spin relaxation causes a nonlinear relation between the oscillator power and the applied voltage bias. 
		Finally, we consider an alternative spin relaxation mechanism and study the resulting magnetization precessions, highlighting the importance of understanding the exact nature of the relaxation in nanomagnets.
		
	\end{abstract}
	\maketitle

	\section{Introduction}
	The discovery of the spin-transfer torque (STT) \cite{slonczewskiCurrentdrivenExcitationMagnetic1996,bergerEmissionSpinWaves1996}, where a voltage bias over a magnetic layered structure results in a spin-polarized current, is one of the cornerstones in the field of spintronics . The quasiclassical description of such systems predicts the existence of steady-state precessions of the magnetization due to the STT, creating a spin-torque oscillator \cite{slavinNonlinearAutoOscillatorTheory2009,kiselevMicrowaveOscillationsNanomagnet2003,rippardDirectCurrentInducedDynamics2004}. The phase diagram of these systems is very rich, with features such as stochastic switching of the magnetization, small and large-amplitude precessions and more \cite{katineCurrentDrivenMagnetizationReversal2000,grollierSwitchingSpinValve2003,kiselevMicrowaveOscillationsNanomagnet2003}. Understanding the full breadth of the phase diagram is usually done within the classical framework of the Landau-Lifshitz-Gilbert-Slonczweski (LLGS) equation \cite{lakshmananFascinatingWorldLandau2011}. However, achieving quantitative and qualitative agreement with spin-torque oscillator experiments is not always possible \cite{xiaoMacrospinModelsSpin2005,berkovMagnetizationPrecessionDue2005,mistralCurrentdrivenMicrowaveOscillations2006}. To resolve this conflict, multiple solutions have been proposed, such as modifying the Gilbert damping part of the LLG equation \cite{tiberkevichNonlinearPhenomenologicalModel2007}, disregarding the macrospin assumption \cite{berkovTransitionMacrospinChaotic2005} and others \cite{berkovSpintorqueDrivenMagnetization2008}.

	\begin{figure}
		\includegraphics[width=\columnwidth]{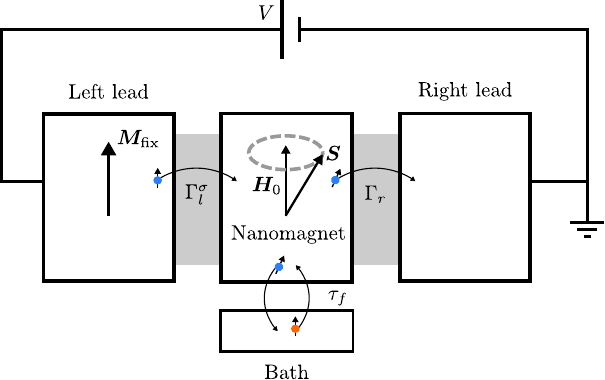}
		\caption{The system considered in this work: a zero-dimensional nanomagnet, subject to an external magnetic field, tunnel coupled to two leads. The left lead is magnetic with a fixed direction of magnetization, while the right lead is a normal metal. To model spin relaxation processes, the nanomagnet is allowed to interact with a reservoir, modeling a fictitious lead (indicated as bath).
			In the presence of finite (non-infinite) relaxation on the nanomagnet, the applied voltage bias and the precessing magnetization drive the electron system away from equilibrium; thus, necessitating a full nonequilibrium description.
			\label{fig:system}}
	\end{figure}
	
	From a fundamental as well as a practical point of view, the existence of dynamical precessions of the magnetization as a voltage is applied is of particular interest.
	Because of the interaction between the magnetization and electron spins, the precessing magnetization drives the electron system away from equilibrium.
	However, conventionally the true nonequilibrium nature of the system is not taken into account, and the Slonczweski STT is commonly derived by assuming an equilibrium distribution function for the electrons on the nanomagnet, neglecting potentially important nonequilibrium effects \cite{tserkovnyakTunnelbarrierenhancedDcVoltage2008}. In this work, in contrast, we offer a full nonequilibrium treatment of the system, in order to explicitly capture the out-of-equilibrium nature of the dynamical precessions.
	
	We focus on including dissipation through spin relaxation processes, motivated by two of our previous works  \cite{ludwigStrongNonequilibriumEffects2017,gunninkChargeConservationSpintorque2024}. In \textcite{ludwigStrongNonequilibriumEffects2017} we found that in the absence of relaxation processes a self-induced torque exists, which is however of such a form that steady state precessions are no longer allowed. 
	In \textcite{gunninkChargeConservationSpintorque2024} we considered the opposite limit, by forcing an equilibrium distribution on the electrons---motivated by the fact that in the presence of strong internal relaxation such a distribution should exist. We showed that imposing charge conservation in conjunction with the LLGS equation also gives rise to a similar self-induced torque. However, this self-induced torque is now of such a form that steady-state precessions are allowed. In addition, the self-induced torque gave qualitative corrections to the conventional LLGS description.

	Our previous works \cite{ludwigStrongNonequilibriumEffects2017,gunninkChargeConservationSpintorque2024} have thus demonstrated that the competition between nonequilibrium effects and (internal) relaxation processes is critically important for a proper description of magnetization dynamics in STT systems. Other works have come to a similar conclusion \cite{hammarTimedependentSpinTransport2016,hammarTransientSpinDynamics2017,bajpaiSpintronicsMeetsNonadiabatic2020, bajpaiTimeretardedDampingMagnetic2019,chenSpinChargePumping2009,petrovicSpinChargePumping2018}.

	In this work we therefore develop a formalism that allows us to treat the relaxation rates on an equal footing with the tunnel coupling to the leads. This gives us access to the magnetization dynamics in the intermediate regime, where the internal relaxation processes are comparable to the tunneling from and to the leads. 
	We employ the nonequilibrium Keldysh formalism to study the system, since the precessions are dynamically stabilized and therefore are, by definition, out of equilibrium. 
	To model internal relaxation processes, we allow the interaction of the nanomagnet’s electron system with a \emph{fictitious} lead; in the following, we also refer to the fictitious lead simply as a \emph{bath}.

	This article is outlined as follows. In \cref{sec:setup} we describe the setup considered, including how we intend to treat the relaxation processes. \Cref{sec:method} forms the main derivation of the work, where we give a full nonequilibrium description of the system. This section is self-contained and can be skipped, since in \cref{sec:stationary-solutions} we summarize the results of the previous section, before moving on to the resulting quasiclassical magnetization dynamics. In \cref{sec:anisotropy} we consider the magnetization dynamics in the presence of magnetic anisotropy, which most closely represents experimental setups. Next, in \cref{sec:rotating} we explore the possibility of an alternative relaxation process which we introduce in a frame rotating with the preccesing magnetization and discuss the resulting quasiclassical dynamics. We end with a conclusion and discussion in \cref{sec:conclusion}. In \cref{app:deltaQ,app:stability,app:definitions} we give further details of the calculations and a list of definitions used throughout this work.

	\section{Setup}
	\label{sec:setup}
	We consider a metallic nanomagnet, subject to an external magnetic field $\bm{H}_0$ and tunnel coupled to two leads, see Fig.~\ref{fig:system}. We will assume here that the magnetization is formed by localized magnetic moments\footnote{Alternatively, one could assume that the Stoner mechanism is the primary driver of the magnetism, which can be treated in a similar way leading to analogous results. See for example Ref.~\cite{ludwigStrongNonequilibriumEffects2017}.} and can be described by a macrospin $\bm S(t)$, which couples to the itinerant electrons on the nanomagnet through an sd-like coupling. The left lead is an itinerant ferromagnet with fixed magnetization $\bm M_{\mathrm{fix}}$, while the right lead is a normal metal. Applying a voltage bias between the left and right lead will drive a spin-polarized current across the left tunnel junction, which can potentially drive the macrospin into precession around the external field $\bm H_0$, thus realizing a spin-torque oscillator. 
	Furthermore, the leads allow for (external) relaxation of angular momentum on the nanomagnet by exchanging electrons with the leads. To model internal angular-momentum relaxation on the nanomagnet, which is the main focus of this work, we also include an additional fictitious lead as a bath.
	This bath is chosen such that it relaxes the electron system to thermal equilibrium in the laboratory frame. 
	
	It is worthwhile to first discuss the nature of equilibrium in this system and the role the leads play. Initially, as no voltage is applied over the system, the itinerant electrons on the nanomagnet are in thermal equilibrium with their environment and all distribution functions follow the Fermi-Dirac (FD) distribution. More specifically, there is a heat bath formed by the two real leads and the fictitious lead, with which the itinerant electrons on the nanomagnet are in thermal equilibrium.
	If a voltage is applied---in particular if the macrospin is driven into a steady-state precession---the nanomagnet’s electron system is then out of equilibrium with its bath. It will therefore exchange energy (and possibly spin) with the two leads and the bath. Importantly, only the distribution functions of the two real leads and the bath are specified and thus follow a Fermi-Dirac distribution.
	The electrons on the nanomagnet do not follow a Fermi-Dirac distribution. Instead they follow a complicated nonequilibrium distribution, as they are driven away from equilibrium by the applied voltage and the precessing macrospin but towards equilibrium by the internal relaxation modeled with the fictitious lead. It is for this reason that one cannot apply the usual assumption of spin pumping in equilibrium magnetic tunnel junctions \cite{tserkovnyakTunnelbarrierenhancedDcVoltage2008} and has to treat the system fully out of equilibrium.
	
	The main goal of this work is to incorporate the internal relaxation processes within the nonequilibrium framework. We thus introduce a fictitious lead, which acts as a bath for the itinerant electrons on the nanomagnet. Such a semi-phenomenological description of relaxation was first introduced by \textcite{buttikerRoleQuantumCoherence1986}. The interactions between the system and a fictitious lead represent the relaxation processes. 
	The lead does not really exist, but it is introduced theoretically to represent internal relaxation mechanisms. Therefore, the fictitious lead cannot function as an electron sink or source, i.e., we require that the average charge current from the nanomagnet to the fictitious lead vanishes, such that it conserves the electron number. However, as we do not assume spin to be conserved, there is no requirement that the spin current from the nanomagnet to the fictitious lead vanishes and thus the fictitious lead can act as a simple model for a spin sink.

	The dynamics of the magnetization, in absence of interactions with the itinerant electron system, are described by
	\begin{equation}
		\partial_t \bm{s} = \bm{s} \cross \bm{H}_{\mathrm{eff}},
		\label{eq:LLG}
	\end{equation}
	where $\bm s = \bm S / S$ is the direction of the macrospin $\bm{S}=S\left(\sin\theta\cos\phi, \sin\theta\sin\phi, \cos\theta\right)$ of length S. The spin precesses around an effective magnetic field $\bm{H}_{\mathrm{eff}}\equiv H_{0} +2 KS_z$, where we have allowed for an easy-axis anisotropy with strength $K>0$.
	
	We describe the electrons on the nanomagnet with the Hamiltonian
	\begin{equation}
		H_e=\Hdot + H_l+H_r+H_{\bath}+H_{t}.
	\end{equation}
	The itinerant electrons on the nanomagnet are described by the Hamiltonian 
	\begin{equation}
		\Hdot =\sum_{\alpha\sigma}\varepsilon_{\alpha} c_{\alpha\sigma}^\dagger c_{\alpha\sigma} - \frac{J}{2}\sum_{\alpha\sigma\sigma'} \hat{\bm{S}} \cdot \bm{\sigma}_{\sigma\sigma'} c_{\alpha\sigma}^\dagger c_{\alpha\sigma'},
	\end{equation}
	where $\varepsilon_{\alpha}$ are the non-interacting single-particle energies and $(J/2)\hat{\bm{S}} \cdot \bm{\sigma}_{\sigma\sigma'}$ describes the coupling between electrons with spins $\sigma, \sigma'$ and the macrospin $\hat{\bm S}$, with coupling strength $J$.

	As is shown in Fig.~\ref{fig:system}, the nanomagnet is coupled to three different leads, including a fictitious lead which represents relaxation. The left, right, and fictitious lead are described by 
	\begin{align}
		H_l &= \sum_{n\in N_l} \sum_\sigma\int\frac{\dd{k}}{2\pi} (\varepsilon_{nk}-M_{\mathrm{fix}}\sigma/2+\mu_l)a_{nk,\sigma}^\dagger a_{nk,\sigma},\\
		H_r &= \sum_{n\in N_r}\sum_\sigma \int\frac{\dd{k}}{2\pi} (\varepsilon_{nk}+\mu_r)a_{nk,\sigma}^\dagger a_{nk,\sigma}, \\
		H_{\bath} &= \sum_{n\in N_{\bath}}^{N_r}\sum_\sigma \int\frac{\dd{k}}{2\pi} (\varepsilon_{nk}+\mu_{\bath})a_{nk,\sigma}^\dagger a_{nk,\sigma},
	\end{align}
	where $-M_{\mathrm{fix}}\sigma/2$ accounts for the different energy of spin up versus spin down electrons in the magnetized left lead and $\varepsilon_{nk}$ are the single particle energies of the left, right and fictitious lead. Here, $\mu_{l/r}$ is the electrochemical potential of the left/right lead, which is controlled externally. The electrochemical potential of the fictitious lead, $\mu_{\bath}$, will be fixed by requiring that charge is conserved in the interaction with the fictitious lead.
	The tunnel couplings between the nanomagnet and the leads are given by 
	\begin{equation}
		H_{t} = \sum_{n} \sum_{\alpha\sigma} \int\frac{\dd{k}}{2\pi} t_{\alpha n} c_{\alpha\sigma}^\dagger a_{nk,\sigma} + \mathrm{H.c.}\,,
	\end{equation}
	where $t_{\alpha n}$ is the tunneling amplitude. 
	
	The distribution function of the three leads all follow the Fermi-Dirac distributions, as we assume the leads to be in (local) equilibrium. For simplicity, we assume the leads to all have the same temperature $T$. We allow for a voltage bias over the nanomagnet, i.e., $\mu_r=\mu$ and $\mu_l=\mu+V$, driving the system out of equilibrium. 
	
	In what follows, we will describe our formalism to account for the nonequilibrium nature of the dynamical precession. This section can be skipped if desired, as we give the main result, the quasiclassical equation of motion describing the magnetization dynamics as a function of voltage, in \cref{sec:stationary-solutions}.

	\section{Method}
	\label{sec:method}
	Since the system we consider here is not in equilibrium, we use the Keldysh formalism, in the path integral form \cite{altlandCondensedMatterField2010,kamenevFieldTheoryNonEquilibrium2011}. The Keldysh generating function is given by
	\begin{equation}
		\mathcal{Z}=\int D\left[\bar{\Psi},\Psi,g\right] e^{i\mathcal{S}_\Psi+ i \mathcal{S}_g},
	\end{equation}
	where $\mathcal{S}_\Psi \left[\bar{\Psi},\Psi,g\right]$ is the electron action and $\mathcal{S}_g[g]$ is the macrospin action. Here $\bar{\Psi},\Psi$ denote fermionic fields and $|g\rangle$ denote spin coherent states \cite{altlandCondensedMatterField2010}. The actions are given by
	\begin{align}
		\mathcal{S}_{\Psi} & =\ii\oint_K \dd{t} \left[ \bar{\Psi}i\partial_{t}\Psi- H\left(\bar{\Psi},\Psi,g\right) \right],\\
		\mathcal{S}_{g} & =\ii\oint_K \dd{t} \ \langle g|i\partial_{t}-\bm{H}_0\cdot \hat{\bm{S}}-K \hat S_z^2|g\rangle \label{eq:Sg},
	\end{align}
	where $\hat{\bm{S}}$ is a vector of spin operators in the spin-$S$ representation and $K$ is the uniaxial magnetocrystalline anisotropy constant. Note that with the spin coherent states, $\bm S=\bra{g}\hat{\bm{S}}\ket{g}$.
	
	\subsection{Electron dynamics}
	The fields of the leads can be integrated out, since they only enter up to quadratic order, such that we obtain
	\begin{equation}
		\mathcal{S}_{\Psi}  =\ii\oint_K \dd{t} \left[ \bar{\Psi}\left(i\partial_{t}-\hat{\Sigma}\right)\Psi- H\left(\bar{\Psi},\Psi\right) \right], \label{eq:SPsi-trln}
	\end{equation}
	where $\hat{\Sigma}\Psi(t)=\oint_K \dd{t'} \Sigma(t-t')\Psi(t')$ and $\Sigma=\Sigma_l+\Sigma_r+\Sigma_{\bath}$ is the self-energy related to tunneling between the leads and the nanomagnet, given by $\Sigma_{l/r/f}=t_{l/r/f}G_{l/r/f}t^\dagger_{l/r/f}$. 
	Within the Keldysh formalism the self-energies carry information about both the level broadening of the states on the nanomagnet and the distribution function of the leads. The nanomagnet does not have a distribution function of its own and instead the distribution of the electrons on the nanomagnet is enslaved to the distribution function of the leads, in combination with the dynamics of the nanomagnet. 
	
	We assume many weakly and randomly coupled transport channels, which allows us to approximate the tunneling by just three tunneling rates: $\Gamma_l^\uparrow$ and $\Gamma_l^\downarrow$ for the spin-dependent coupling to the left lead and $\Gamma_r$ for the coupling to the right lead \cite{ludwigStrongNonequilibriumEffects2017}.\footnote{Note that the superscript refers here to the spin state in the leads.} The interaction with the bath described by $H_{\bath}$ gives rise to a relaxation rate $\tau_{\bath}^{-1}$.\footnote{The relaxation rate $\tau_f^{-1}$ corresponds to a level broadening or tunneling rate $\Gamma_f=\tau_f^{-1}$ for tunneling to the fictitious lead. However, to stress the difference between real and fictitious leads in the notation, we use $\tau_f^{-1}$ instead of $\Gamma_f$.} Here it is important to note that the lead between the nanomagnet and the bath is a fictitious lead which represents spin relaxation, and therefore its electrochemical potential $\mu_{\bath}$ will be determined self-consistently to ensure that no charge current flows between the nanomagnet and the fictitious lead.
	
	The macrospin dynamics are thus governed by the electron action given in \cref{eq:SPsi-trln}, in combination with the macrospin action \cref{eq:Sg}. The macrospin action, $\mathcal{S}_g$, already contains only the spin coherent states and therefore its contribution to the macrospin dynamics can be readily read off, and will result in \cref{eq:LLG}. It is however coupled to the electron action, $\mathcal{S}_{\Psi}$, through the sd-like coupling. 
	To find the effective macrospin dynamics, which we are interested in, we integrate out the electron system. 
	
	In the following subsections we perform this integrating of the electron system. In order to achieve this, we introduce the rotating frame, expand the action in quantum components, derive the classical Green's functions and the electrochemical potential of the fictitious lead, and finally derive a full quasiclassical action.

	\subsection{Rotating frame}
	\label{sec:rotating-frame}
	The macrospin $\bm{S}(t)$ is precessing and the action $i\mathcal{S}_{\Psi}$ in \cref{eq:SPsi-trln} is thus non-trivial. We therefore apply a rotation into a frame where $\bm{S}$ is always pointing along the $z$ axis, i.e. $R^\dagger \bm{S}\cdot\bm{\sigma}R=S\sigma_z$. However, since the macrospin direction will depend on time, the rotation $R$ will also depend on time, introducing an extra term $Q=-\ii R^\dagger \dot{R}$ in the action. The final action is then, after integrating out the fermionic fields and re-exponentiating,
	\begin{equation}
		\ii\mathcal{S}_{\Psi} = \Tr\ln\left[G_0^{-1} + JS\frac{\sigma^z}{2}-R^\dagger\Sigma R- Q\right].\label{eq:iS-rotating}
	\end{equation}
	Here $G_0^{-1}\equiv i\partial_t - \varepsilon_\alpha $.
	For the rotation matrix $R$ we choose the Euler angle representation, 
	$R=e^{-i\phi\sigma_z/2}e^{-\ii\theta\sigma_y/2} e^{\ii(\phi-\chi)\sigma_z/2}$,
	where $\chi$ is a gauge freedom, and $\theta, \phi$ describe the direction of the macrospin (in the lab frame), i.e., $\bm S = S(\sin \theta \cos \phi, \sin \theta \sin \phi, \cos \theta)$ \cite{shnirmanGeometricQuantumNoise2015}. 
	
	We use the gauge freedom to choose $\dot{\chi}_c=\dot{\phi}_c(1-\cos\theta_c)$ and ${\chi}_q={\phi}_q(1-\cos\theta_c)$, which are chosen such that they will eliminate certain parts of the action \cite{shnirmanGeometricQuantumNoise2015,ludwigThermallyDrivenSpin2019}; as shown in \cref{app:deltaQ}. 
	
	We now assume that the change of macrospin dynamics is slow compared to the time scales of the individual electron spins, i.e., we assume that the electrons follow the macrospin adiabatically. 
	
	\subsection{Expansion in quantum components}
	\label{sec:expand}
	We are interested in the quasiclassical equations of motion, which can be obtained from varying the action with respect to the quantum components. We therefore perform the standard rotation of the Keldysh components to the $(c,q)$ basis and split the rotations into two parts, one purely classical, defined by $R_k\equiv R_c|_{q=0}$ and a remainder $\delta R \equiv R - R_k$ \cite{ludwigThermallyDrivenSpin2019}. Note that $R_c\equiv (R_+ + R_-)/2\neq R_k$ if the quantum components of the dynamical variables do not vanish. We proceed analogously for $\delta \tilde{Q} \equiv \tilde{Q} - \tilde{Q}_{k}$.

	We now write 
	\begin{equation}
		\ii \mathcal{S}_{\Psi} = \Tr\ln\left[ \tilde{G}_c^{-1} - \delta\Sigma -\delta \tilde{Q}\right],
	\end{equation}
	where $\tilde{G}_{c}^{-1}=G_0^{-1}+JS^{z}\frac{\sigma^{z}}{2}-Q_{k}-R^{\dagger}_k\Sigma R_{k}$ and  $\delta\Sigma = R^\dagger \Sigma R - R_k^\dagger \Sigma R_{k}$. The classical Green's function $G_c$ thus contains no quantum components, which are all shifted into $\delta \tilde{Q}$ and $\delta\Sigma$, which are by definition of first and higher (but not zeroth) order in quantum components.
	
	We now expand the action in quantum components, and we obtain two contributions to the electron action:
	\begin{align}
		\ii\mathcal{S}_{Q} &= -\Tr\left[\tilde{G}_c \delta\tilde{Q} \right], \label{eq:S_Q}	\\
		\ii\mathcal{S}_{\mathrm{AES}} &= -\Tr\left[\tilde{G}_c \delta\Sigma\right]. \label{eq:AES}
	\end{align}
	The first contribution, containing the parts of the actions with $\delta\tilde{Q}$, turns out to be irrelevant for the resulting macrospin dynamics. We show this in more detail in \cref{app:deltaQ}. Thus, the remaining contribution to the electron action is $\mathcal{S}_{\mathrm{AES}}$, which is an Ambegaokar-Eckern-Schön (AES) type action \cite{ambegaokarQuantumDynamicsTunneling1982,eckernQuantumDynamicsSuperconducting1984,shnirmanGeometricQuantumNoise2015}. It carries information regarding the tunneling of electrons between the leads and the nanomagnet, and thus describes the macrospin dynamics, in combination with the macrospin action $\mathcal{S}_g$.
	
	We perform the same procedure for the macrospin action $\mathcal{S}_g$, expanding up to first order in the quantum components to obtain
	\begin{equation}
		\mathcal{S}_{g}  =\ii\oint_K \dd{t} \ S \left[
		\theta_q \sin\theta_c \left(-H_0-2KS\cos\theta_c + \dot\phi_c\right) - \phi_q\sin\theta_c \dot\theta_c \right] \label{eq:Sg-expanded}.
	\end{equation}
	
	\subsection{Classical Green's function}
	\label{sec:classical-green}
	Next, we need to determine the classical Green's function $\tilde{G}_c$. We assume $JS$ to be large, and thus $\tilde{G}_c$ is approximately diagonal, with the off-diagonal parts suppressed by $1/JS$. Its inverse is then given by $\tilde{G}_c^{-1}=G_0^{-1}+JS^z\sigma^z/2-R_k^\dagger\Sigma R_{k}$. We use the slowness of the classical coordinates $\dot\phi_c,\theta_c$ to determine the rotating self-energy approximately and find (see Appendix B in Ludwig \textit{et al.} \cite{ludwigThermallyDrivenSpin2019} for further details)
	\begin{align}
		\tilde{G}_{s}^{R/A}\left(\bar{t},\omega\right) & =\frac{1}{\omega-\xi_{\alpha\sigma}\pm \ii\Gamma_{\sigma}\left(\theta_{c}\right)}\label{eq:GRA}\\
		\tilde{G}_{s}^{K}\left(\bar{t},\omega\right) & =\frac{-2i\Gamma_{\sigma}\left(\theta_{c}\right)}{\left(\omega-\xi_{\alpha\sigma}\right)^{2}+\Gamma_{\sigma}^{2}\left(\theta_{c}\right)}\tilde F_{s}^{\sigma}\left(\bar{t},\omega\right)\label{eq:GK}
	\end{align}
	where we introduced $\bar{t}=\frac{t+t'}{2}$ and $\tilde{t}=t-t'$ (the Wigner coordinates) and performed the Fourier transform with respect to $\tilde{t}$. Here  $\xi_{\alpha\sigma}=\epsilon_\alpha-JS\sigma/2$ denote the single-particle energy for level $\alpha$ and spin $\sigma$ and we have introduced the level broadenings $\Gamma_l^\sigma(\theta)\equiv \cos^2\frac{\theta}{2} \Gamma_l^\sigma + \sin^2\frac{\theta}{2} \Gamma_l^{\bar{\sigma }}$ and $\Gamma_\sigma\left(\theta\right)\equiv\Gamma_l^\sigma(\theta)+ \Gamma_r +\tau_{\bath}^{-1}$. Here $\Gamma_{\uparrow/\downarrow}^l$ is the level broadening due to interaction with the left lead for spin $\uparrow$/$\downarrow$, $\Gamma_r$ and $\tau^{-1}_{\bath}$ are the level broadening due to interaction with the right and fictitious leads, which are  both spin-independent.  The nanomagnet's dynamical but slow distribution function is given by 
	\begin{equation}
		\begin{split}
			\tilde F_{s}^{\sigma}\left(\bar{t},\omega\right)=&\frac{1}{\Gamma_{\sigma}\left(\theta_{c}\right)} \Bigr[\cos^{2}\frac{\theta_{c}}{2}\Gamma_{l}^{\sigma}F_{l}\left(\omega+\sigma\omega_{-}\right) \\
			&+\sin^{2}\frac{\theta_{c}}{2}\Gamma_{l}^{\bar{\sigma}}F_{l}\left(\omega+\bar{\sigma}\omega_{+}\right)\\
			&+\cos^{2}\frac{\theta_{c}}{2}\Gamma_{r}F_{r}\left(\omega+\sigma\omega_{-}\right)\\
			&+\sin^{2}\frac{\theta_{c}}{2}\Gamma_{r}F_{r}\left(\omega+\bar{\sigma}\omega_{+}\right)\\
			&+\cos^{2}\frac{\theta_{c}}{2}{\tau_{\bath}^{-1}} F_{\bath}\left(\omega+\sigma\omega_{-}\right)\\
			&+\sin^{2}\frac{\theta_{c}}{2}{\tau_{\bath}^{-1}} F_{\bath}\left(\omega+\bar{\sigma}\omega_{+}\right)
			\Bigr],
		\end{split}
		\label{eq:slow-dist-function}
	\end{equation}
	where $F_{\eta} = \tanh[\frac{\omega-\mu_{\eta}}{2T_{\eta}}]$ are the leads' distribution functions with $\eta\in\{l,r,f\}$ and the Berry phase enters via the dynamic shifts $\omega_{\pm}=\dot{\phi}_c\left[1\pm\cos\theta_c\right]/2$. From Eq.~(\ref{eq:slow-dist-function}) it is clear that the distribution function is a superposition of the distribution functions of the three leads, as well as a function of the magnetization dynamics through the Berry phase $\omega_{\pm}$. The nanomagnet’s electron system is thus manifestly out of equilibrium.

	\subsection{Charge current}
	We derive the charge currents, $I_c^{n\rightarrow \eta}$, between the nanomagnet and the lead $\eta\in\{l,r,\bath\}$ by use of the counting field approach \cite{virtanenSpinPumpingTorque2017}, as in \cite{ [{Appendix A in }] Ludwig2019_1000100744}. As a result, we obtain the Landauer-type formula \cite{landauerSpatialVariationCurrents1957,nazarovQuantumTransportIntroduction2009} for the charge current from the nanomagnet to lead $\eta$ 
	\begin{equation}
		I_c^{n\rightarrow\eta}  =\frac{1}{2}\sum_{\sigma}\int d\omega\ 2\rho_{\sigma}(\omega)\Gamma_{\eta}^\sigma(\theta)\left[\tilde F_{\eta}^\sigma\left(\omega\right)-\tilde{F}_{s}^{\sigma}\left(\omega\right)\right],
		\label{eq:Id3}
	\end{equation}
	where $\rho_{\uparrow/\downarrow}$ is the nanomagnet's  spin-dependent density of states and $\eta\in\{l,r,\bath\}$. Note that this result is valid for any lead, including the fictitious lead. Finally, the charge current through the system is given by $I_c\equiv I_c^{l\rightarrow n} = - I_c^{r\rightarrow n}$, since charge is conserved in the interaction with the fictitious lead.

	\subsection{Fictitious lead}
	\label{sec:fictitious-lead}
	As can directly be seen from Eq.~(\ref{eq:slow-dist-function}), the electrochemical potential of the fictitious lead affects the distribution function on the nanomagnet and, in turn, it also affects the currents flowing into and out of the nanomagnet and the nanomagnet’s magnetization dynamics. Since this is not a real lead, but charge is conserved, its electrochemical potential $\mu_{\bath}$ must be determined such that the charge current between the nanomagnet and the lead vanishes, i.e., $I_{n\rightarrow\bath}=0$ \cite{buttikerRoleQuantumCoherence1986}. However, we allow the spin current to be non-zero, which corresponds to a bath which does not conserve spin. In other words, the nanomagnet can exchange spin with the bath, but not charge.
	
	Now the electrochemical potential of the lead can be found by requiring that $I^{n\rightarrow\bath}_c=0$, where , 
	\begin{multline}
		I_c^{n\rightarrow\bath}=2\sum_\sigma \frac{\rho_\sigma}{\tau_{\bath}\Gamma_\sigma(\theta)}\times\\\left[\Gamma_r (\mu_r-\mu_{\bath})+\Gamma_l^\sigma(\theta)(\mu_l-\mu_{\bath})+\frac12 \Gamma_\Delta \sin^2\theta\,\dot\phi\right]. \label{eq:Ic-bath}
	\end{multline}
	Now solving $I_c^{n\rightarrow\bath}=0$ for the electrochemical potential of the bath $\mu_{\bath}$ we find that
	\begin{widetext}
		\begin{equation}
			\mu_{\bath} = \frac{\rho_\uparrow \Gamma_\downarrow(\theta)\left( \Gamma_r \mu_r + \Gamma_l^\uparrow(\theta) \mu_l + \frac{1}{2}\sin^2\theta\, \Gamma_{\Delta} \dot \phi  \right) + \rho_\downarrow \Gamma_\uparrow(\theta)\left( \Gamma_r \mu_r + \Gamma_l^\downarrow(\theta) \mu_l + \frac{1}{2}\sin^2\theta\, \Gamma_{\Delta} \dot \phi  \right)}{\rho_\downarrow\Gamma_\uparrow(\theta) \left(\Gamma_l^\downarrow(\theta)+\Gamma_r\right)+\rho_\uparrow\Gamma_\downarrow(\theta) \left(\Gamma_l^\uparrow(\theta)+\Gamma_r\right)}.\label{eq:mu3-nolim}
		\end{equation}
	\end{widetext}
	The electrochemical potential $\mu_{\bath}$ obtained in this way can now be reinserted into the distribution function $F_{\bath}(\omega)$ in Eq.~\eqref{eq:slow-dist-function}. Now, the effective magnetization dynamics can be explicitly described.

	\subsection{Quasiclassical action}
	\label{sec:quasiclassical-action}
	As was shown before, the quasiclassical action is dominated by the AES-action, Eq.~(\ref{eq:AES}), which we write as
	\begin{multline}
		\mathcal{S}_{\mathrm{AES}} =-\sum_{\sigma\sigma'}\int\dd{t}\int \dd{t'}\Im\Big[R_{q}^{\sigma\sigma'}\left(t\right)\\
		\times\alpha_{\sigma\sigma'}\left(t,t'\right) \left(R_{c}^{\sigma'\sigma}\left(t'\right)\right)^*\Big],
	\end{multline}
	with the kernel function defined as
	\begin{multline}
		\alpha_{\sigma\sigma'}\left(t,t'\right)=\Tr\big[\tilde{G}_{s\sigma}^{R}\left(t,t'\right)\tilde{\Sigma}_{\sigma'}^{K}\left(t'-t\right)\\
		+\tilde{G}_{s\sigma}^{K}\left(t,t'\right)\tilde{\Sigma}_{\sigma'}^{A}\left(t'-t\right)\big].
	\end{multline}
	More explicitly, we find
	\begin{multline}
		\alpha_{\sigma\sigma'}(\bar{t},\omega) = \int\dd{\omega'} \rho_\sigma(\omega') \Big[ \Gamma_l^{\sigma'} \left[F_s^\sigma(\bar{t},\omega') - F_l(\omega'-\omega)\right] \\
		+\Gamma_r\left[F_s^\sigma(\bar{t},\omega') - F_r(\omega'-\omega)\right]
		\Big].
	\end{multline}
	We split the kernel function based on the frequency dependence, $\alpha_{\sigma\sigma'}(\bar t,\omega) = \alpha^d_{\sigma\sigma'}(\omega)+\alpha^s_{\sigma\sigma'}(\bar{t})$, defining a dissipative contribution $\alpha^d_{\sigma\sigma'}(\omega)=\alpha_{\sigma\sigma'}(\bar{t},\omega)-\alpha_{\sigma\sigma'}(\bar{t},0)$ (which will result in a Gilbert damping enhancement) and a spin-torque term which is independent of frequency $\alpha^s_{\sigma\sigma'}(\bar{t})=\alpha_{\sigma\sigma'}(\bar{t},0)$ (which will result in a Slonczewski-like torque). Performing the frequency integrations we have
	\begin{align}
		\alpha^d_{\sigma\sigma'}(\omega) &= g_{\sigma\sigma'}\omega, \\
		\alpha^s_{\sigma\sigma'}(\bar{t}) &=I_{V}^{\sigma\sigma'}+I_{h}^{\sigma\sigma'}+I_{f}^{\sigma\sigma'},
	\end{align} 
	where $g_{\sigma\sigma'} = 2\rho_\sigma(\Gamma_l^{\sigma'} + \Gamma_r + \tau_{\bath}^{-1})$ and
	\begin{align}
		I_{V}^{\sigma\sigma'}& =\frac{2\Gamma_{r}\Gamma_{\Delta}}{\Gamma_{\sigma}\left(\theta_{c}\right)}\left(\sigma'-\sigma\cos\theta_{c}\right)\rho_{\sigma}\left(\mu_{l}-\mu_{r}\right),\label{eq:IVsigmasigma}\\
		I_{p}^{\sigma\sigma'} & =g_{\sigma\sigma'}\frac{\Gamma_{\Delta}}{2\Gamma_{\sigma}\left(\theta_{c}\right)}\sin^{2}\theta_{c}\,\dot{\phi}_{c},\label{eq:IHsigmasigma}\\
		I_{\bath}^{\sigma\sigma'}& =\frac{1}{\tau_{\bath}}\frac{2\Gamma_{\Delta}}{\Gamma_{\sigma}\left(\theta_{c}\right)}\left(\sigma'-\sigma\cos\theta_{c}\right)\rho_{\sigma}\left(\mu_{l}-\mu_{\bath}\right),\label{eq:I4sigmasigma}
	\end{align}
	where $\Gamma_\Delta=(\Gamma_l^\uparrow - \Gamma_l^\downarrow)/2$.
	We have split these contributions based on their origin. The first contribution is the result of the applied voltage bias, $V\equiv\mu_l-\mu_r$; the second contribution\footnote{We use the subscript $p$ to indicate that this torque is ultimately related to the pumping of electrons into the adjacent leads due to the preccesing magnetization.} is the result of the precessing macrospin pumping spins and charges into the left magnetic lead and thus proportional to $\dot\phi$; the bath contribution is the result of the internal relaxation, or correspondingly, the interaction with the fictitious lead. 
	
	Here we stress that the fictitious lead also contributes to the Gilbert damping enhancement through the dissipative kernel, $\alpha^d_{\sigma\sigma'}(\omega)$. It will act effectively as an ``internal'' Gilbert damping, since $\tau_{\bath}^{-1}$ is an intrinsic parameter of the system at hand. 
	
	
	Now, with the classical Green's function derived, the full action can be worked out. Up to first order in quantum components, we have for the AES-action that (dropping the index c for classical, such that $\theta\equiv \theta_c, \dot{\phi}\equiv \dot{\phi}_c$) 
	\begin{multline}
		\mathcal{S}_{\mathrm{AES}} = -\int \dd{t} \, \Bigr\{ \theta_q \tilde{g}(\theta)\dot{\theta}
		+\phi_q \sin^2\theta \left[ \tilde{g}(\theta)\dot{\phi} + I_s(\theta) \right] \Bigr\}, 
		\label{eq:final-aes}
	\end{multline}
	where 
	\begin{align}
		\tilde{g}(\theta)&=\frac{{g_{\uparrow\uparrow}+g_{\downarrow\downarrow}}}{4}\sin^2\left(\frac{\theta}{2}\right)+\frac{g_{\downarrow\uparrow}+g_{\uparrow\downarrow}}{4}\cos^2\left(\frac{\theta}{2}\right), \label{eq:gtilde} \\
		&= \frac12 \left[\rho_\Sigma\left( \Gamma_{\Sigma} + \tau_{\bath}^{-1}\right) - \rho_\Delta \Gamma_\Delta \cos\theta\right]
	\end{align}
	is the spin conductance \cite{chudnovskiySpinTorqueShotNoise2008} and
	\begin{equation}
		I_s(\theta,\dot{\phi}) = \frac{1}{4}[\alpha^s_{\uparrow\uparrow}-\alpha^s_{\uparrow\downarrow}+\alpha^s_{\downarrow\uparrow}-\alpha^s_{\downarrow\downarrow}] \label{eq:Isummation}
	\end{equation}
	is the spin-torque current driving the macrospin. We now perform the summation over the spin indices as defined in \cref{eq:Isummation} over the three kernel functions in \cref{eq:IVsigmasigma,eq:IHsigmasigma,eq:I4sigmasigma} to obtain three spin-transfer torques, 
	\begin{align}
		\Ib& =\frac{2\Gamma_{\Delta}\Gamma_{r}}{\Gamma_{\uparrow}\left(\theta\right)\Gamma_{\downarrow}\left(\theta\right)}\tilde{g}\left(\theta\right)V \label{eq:Ib-intermediate}\\
		\Ih& =\frac{\Gamma_{\Delta}^{2}}{\Gamma_{\uparrow}\left(\theta\right)\Gamma_{\downarrow}\left(\theta\right)}\tilde{g}\left(\theta\right)\frac{1}{2}\sin^2\theta\,\dot\phi, \label{eq:Ih-intermediate}\\
		I_{\bath}(\theta,\dot{\phi}) & =\frac{2\Gamma_{\Delta}}{\tau_{\bath}\Gamma_{\uparrow}\left(\theta\right)\Gamma_{\downarrow}\left(\theta\right)}\tilde{g}(\theta)\left(\mu_l - \mu_{\bath}\left(\theta,\dot\phi\right) \right) \label{eq:I4-intermediate}
	\end{align}
	which are again split based on their physical origin: (1) $\Ib$ is proportional to the applied voltage bias $V\equiv\mu_l-\mu_r$; (2) $\Ih$ is proportional to $\dot\phi$ and (3) $I_{\bath}(\theta,\dot\phi)$ follows from the interaction with the fictitious lead. Together, the three spin-torque currents determine the total Slonczewski-like spin-torque current driving the macrospin 
	\begin{equation}
		I_s(\theta,\dot\phi) = \Ib + \Ih + I_{\bath}(\theta,\dot{\phi}).
	\end{equation}
	
	In the absence of internal relaxation, we have that $\tau_{\bath}^{-1}=0$ and thus $I_{\bath}(\theta,\dot{\phi})=0$. Importantly, in this limit, $\Ib$ and $\Ih$ correspond to the results obtained in our previous work, Ref.~\cite{ludwigStrongNonequilibriumEffects2017}, allowing us to focus in this work on the macrospin dynamics for $\tau_{\bath}^{-1}\neq0$.
	\subsection{Charge current}
	Having obtained the quasiclassical action, we can now find the charge current $I_c$ through the system by evaluating \cref{eq:Id3} for $\eta=l$. We find that
	\begin{multline}
		I_c=2\Gamma_r \rho_\Sigma \left[1-\frac{2 \tilde g(\theta) +\rho_\Sigma  \tau_{\bath}^{-1}}{2\tau_{\bath}^{-1} g_e(\theta)+\rho_\Sigma \Gamma_\uparrow(\theta)\Gamma_\downarrow(\theta)} \Gamma_r\right] V\\
		+ \frac{2\tilde g(\theta)+\tau_{\bath}^{-1}\rho_\Sigma}{2\tau_{\bath}^{-1} g_e(\theta)+\rho_\Sigma \Gamma_\uparrow(\theta)\Gamma_\downarrow(\theta)}\Gamma_r\Gamma_\Delta\rho_\Sigma  \sin^2\theta\,\dot\phi, \label{eq:Ic}
	\end{multline}
	where $		g_e\left(\theta\right)\equiv\frac{1}{2}\left[\rho_{\Sigma}\Gamma_{\Sigma}+\rho_{\Delta}\Gamma_{\Delta}\cos\theta\right]$ is the electric conductance for the electrons leaving the nanomagnet into the leads. 
	We stress here that $\theta$ is also a function of the applied voltage $V$, since applying a voltage will drive the system out of equilibrium and possibly into steady-state precessions with $\cos\theta\neq\pm1$.
	
	\subsection{Quasiclassical equations of motion}
	\label{sec:qc-eom}
	We now have a complete expression for the action, $i\mathcal{S} = i\mathcal{S}_g +  i\mathcal{S}_{\mathrm AES}$, where the macrospin and AES-action are given by Eqs.~(\ref{eq:Sg-expanded}) and (\ref{eq:final-aes}). We now vary this action in quantum components, i.e., $\frac{\delta S}{\delta \theta_q}\rvert_{\theta_q=0}=0$ and  $\frac{\delta S}{\delta \phi_q} \rvert_{\phi_q=0}=0$ and obtain the quasiclassical equations of motion as
	\begin{align}
		\sin\theta\, \dot{\phi} &= \sin\theta\,\left[H_0 + 2KS\cos\theta \right]+\frac{\tilde{g}(\theta)}{S} \dot{\theta}, \label{eq:LLGphi} \\
		\sin\theta\, \dot{\theta} &= \frac{\sin^2(\theta)}{S} \left[ \tilde{g}(\theta)  \dot{\phi} +  I_s(\theta,\dot\phi)  \right]. \label{eq:LLGtheta}
	\end{align}
	In order to obtain more insight, these equations of motion can be cast in vector form, with the macrospin direction $\bm s$, such that we find the following Landau-Lifshitz-Gilbert-Slonczewski (LLGS) equation,
	\begin{multline}
		\partial_t \bm{s} = \bm{s} \cross \bm{H}_{\mathrm{eff}} -\alpha\left(\theta\right)\bm{s} \cross \dot{\bm{s}}
		+ \frac{1}{S} I_s(\theta,\dot\phi)\, \bm{s}\cross\left(\hat{\bm{z}}\cross\bm{s}\right),
		\label{eq:LLGS}
	\end{multline}
	where $\alpha(\theta)=(\tilde{g}(\theta+\alpha_0)/S$ and $\bm H_{\mathrm{eff}}\equiv H_0+2KS_z$. Here we have included a phenomenological bulk Gilbert damping $\alpha_0$, which parameterizes the relaxation of the magnetization due to bulk processes.
	There are two main effects here that stem from the interaction with the leads: (1) the Gilbert damping enhancement $\tilde g(\theta)/S$, and (2) a spin transfer torque acting on the macrospin with coefficient $I_s(\theta,\dot\phi)$.

	Formally, we have now completely described the dynamics through the LLGS Equation~\eqref{eq:LLGS}. Importantly, the spin-torque current $I_s(\theta,\dot\phi)$ depends on the interaction with the fictitious lead through its electrochemical potential, Eq.~\eqref{eq:mu3-nolim}. We now continue to study the resulting macrospin dynamics for increasing internal relaxation rates $\tau_{\bath}^{-1}$.

	\section{Results}
	\label{sec:stationary-solutions}
	In the previous section, we have obtained the full quasiclassical equation of motion of the macrospin direction $\bm s = \bm S / S$, where $\bm{S}=S\left(\sin\theta\cos\phi, \sin\theta\sin\phi, \cos\theta\right)$. Here, we summarize these results and find the macrospin dynamics as a voltage bias is applied. 
	
	The macrospin dynamics is described by the Landau-Lifshitz-Gilbert-Slonczewski equation [\cref{eq:LLGS}], 
	\begin{multline}
		\partial_t \bm{s} = \bm{s} \cross \bm{H}_{\mathrm{eff}} -\alpha\left(\theta\right)\bm{s} \cross \dot{\bm{s}}
		+ \frac{1}{S} I_s(\theta,\dot\phi)\, \bm{s}\cross\left(\hat{\bm{z}}\cross\bm{s}\right).
		\label{eq:LLGS2}
	\end{multline}
	Here $\bm H_{\mathrm{eff}}\equiv H_0+2KS_z$ is the effective magnetic field and $\alpha(\theta)=(\alpha_0 + \tilde g(\theta))/S$, where 
	\begin{equation}
		\tilde g(\theta) = \frac12 \left[\rho_\Sigma\left( \Gamma_{\Sigma} + \tau_{\bath}^{-1}\right) - \rho_\Delta \Gamma_\Delta \cos\theta\right]\label{eq:gtilde-final}
	\end{equation} is the Gilbert damping enhancement and $\alpha_0$ is a phenomenological bulk Gilbert damping, which parameterizes the relaxation of the magnetization due to bulk processes.\footnote{Within our current formalism, a bulk Gilbert damping could also be derived through a Caldeira-Leggett approach \cite{verstratenFractionalLandauLifshitzGilbertEquation2023}.} Furthermore,
	\begin{equation}
		I_s(\theta,\dot\phi) \equiv I_V(\theta) + I_p(\theta,\dot\phi) + I_{\bath}(\theta,\dot\phi),
	\end{equation} 
	is the spin-torque current driving the magnetization, which we have split in three possible contributions [\cref{eq:Ib-intermediate,eq:Ih-intermediate,eq:I4-intermediate}], 
	\begin{align}
		\Ib& =\frac{2\Gamma_{\Delta}\Gamma_{r}}{\Gamma_{\uparrow}\left(\theta\right)\Gamma_{\downarrow}\left(\theta\right)}\tilde{g}\left(\theta\right)V, \label{eq:results-sec:Ib-intermediate}\\
		\Ih& =\frac{\Gamma_{\Delta}^{2}}{\Gamma_{\uparrow}\left(\theta\right)\Gamma_{\downarrow}\left(\theta\right)}\tilde{g}\left(\theta\right)\frac{1}{2}\sin^2\theta\,\dot\phi, \label{eq:results-sec:Ih-intermediate}\\
		I_{\bath}(\theta,\dot{\phi}) & =\frac{2\Gamma_{\Delta}}{\tau_{\bath}\Gamma_{\uparrow}\left(\theta\right)\Gamma_{\downarrow}\left(\theta\right)}\tilde{g}(\theta)\left(\mu_l - \mu_{\bath}\left(\theta,\dot\phi\right) \right). \label{eq:results-sec:I4-intermediate}
	\end{align}
	Here, $\tau_{\bath}^{-1}$ is the spin relaxation rate following from interaction with the fictitious lead, $\mu_{\bath}(\theta,\dot\phi)$ is the chemical potential of the fictitious lead [defined in \cref{eq:mu3-nolim}], which is determined self-consistently by imposing charge conservation.
	Furthermore, $\Gamma_\sigma\left(\theta\right)=\cos^2\frac{\theta}{2} \Gamma_l^\sigma + \sin^2\frac{\theta}{2} \Gamma_l^{\bar{\sigma }}+ \Gamma_r +\tau_{\bath}^{-1}$ are the spin-dependent tunneling rates, which importantly are  dependent on $\tau_{\bath}^{-1}$.  Here, $\Gamma_{l,r}^\sigma$ are the spin-dependent level broadenings to the left and right lead, where $\Gamma_r^\uparrow=\Gamma_r^\downarrow\equiv\Gamma_r$. Finally, we have defined $\Gamma_\Delta\equiv\Gamma_l^\uparrow-\Gamma_l^\downarrow$.
	We also summarize these constants in \cref{app:definitions}.
	
	We now discuss the three contributions to the spin-transfer torque $I_s(\theta,\dot\phi)$. First, $I_V(\theta)$ gives rise to a Slonczewski-like torque, proportional to the applied bias $V\equiv\mu_l-\mu_r$. It was first predicted in 1996 \cite{slonczewskiCurrentdrivenExcitationMagnetic1996,bergerEmissionSpinWaves1996} and has been used to successfully explain a wide range of experimental results \cite{kiselevMicrowaveOscillationsNanomagnet2003,berkovMagnetizationPrecessionDue2005,kudoMeasurementNonlinearFrequency2009,slavinNonlinearAutoOscillatorTheory2009}. Most importantly, it can result in precession of the macrospin.  
	
	Second, $I_p(\theta,\dot\phi)$ is proportional to the precession frequency $\dot\phi$, which we refer to as ``self-induced'', because it is only non-zero if $\dot\phi\neq0$, i.e., if the macrospin is precessing. Physically, it originates from the precessing magnetization \emph{pumping} spin into the left lead, which is why we use the subscript $p$. 
	It has the same direction as the Slonczewski torque, but is dependent on both $\theta$ and $\dot{\phi}$. 
	We have first obtained the self-induced torque in Ref.~\cite{ludwigStrongNonequilibriumEffects2017} by considering a nonequilibrium nanomagnet in the absence of internal relaxation.\footnote{In Ref.~\cite{ludwigStrongNonequilibriumEffects2017}, the self-induced torque is referred to as a ``hybrid'' torque.} Within our current formalism we can reproduce these results, such that Eq.~\eqref{eq:results-sec:Ih-intermediate} in our work reduces to Eq.~(44) in Ref.~\cite{ludwigStrongNonequilibriumEffects2017} when we set $\tau_{\bath}^{-1}=0$. 
	In addition, a similar, although not identical, effect can be derived by imposing charge conservation in combination with a conventional LLGS equation, as we have done in Ref.~\cite{gunninkChargeConservationSpintorque2024}, cf. Eq.~\eqref{eq:results-sec:Ih-intermediate} in our work and Eq.~(10) in Ref.~\cite{gunninkChargeConservationSpintorque2024}. However, we stress here that our current formalism does not reproduce this result directly. We will return to this issue in \cref{sec:rotating}, where we propose a simple modification to our formalism to reproduce the results of Ref.~\cite{gunninkChargeConservationSpintorque2024}.
	
	Finally, $I_{\bath}(\theta,\dot\phi)$ is the result of the interaction with the bath. It is dependent on the chemical potential $\mu_{\bath}(\theta,\dot\phi)$ of the fictitious lead, defined in \cref{eq:mu3-nolim}. It is therefore highly non-trivial, but it has an important effect on the macrospin dynamics, which we will consider next.
	
	Importantly, all three contributions to the spin-transfer torque are strongly dependent on the relaxation rate $\tau_{\bath}^{-1}$, and we will show that the relaxation rate quantitatively and qualitatively changes the macrospin dynamics. 	In what follows we have chosen $\rho_{\Sigma}\times H_0=0.2$,
	$\Gamma_{\Sigma}/H_0=0.2$,
	$\rho_{\Delta}\times H_0 =0.1$,
	$\Gamma_{\Delta}/H_0=-0.1$
	, $\Gamma_r/H_0=0.05$. 
	These values have been chosen for demonstration purposes and do not necessarily correspond to a specific experimental realization.

	For clarity, we first assume that anisotropy and the bulk Gilbert damping are absent, i.e., $K=0$ and $\alpha_0=0$. The steady state solutions can then be found by using the ansatz $\phi=\omega_{0} t+\delta\phi$ and $\theta=\theta_0+\delta\theta$. We immediately obtain $\omega_0=H_0+ O(1/S^2)$ and we can solve for $\theta_0$. 
	The stability of these solutions can be determined from the dynamics of the fluctuations, described by $\delta\dot\theta=C_0 \delta\theta$, such that if $C_0<0$, the corresponding $\theta_0$ is stable. We relegate a further derivation to \cref{app:stability} and will simply indicate the resulting stability in the figures.

	\subsection{No internal relaxation}
	\label{sec:no-internal-relaxation}
	In the absence of internal relaxations, i.e., for $\tau_{\bath}^{-1}=0$, we have that $I_{\bath}(\theta,\dot\phi)=0$. This is the regime that was previously studied in Ref.~\cite{ludwigStrongNonequilibriumEffects2017}, and there are no stable precessions allowed. Note that this changes in the presence of anisotropy, as we will discuss in \cref{sec:anisotropy}.

	Specifically, there are no precessions allowed because the self-induced torque perfectly cancels the $\theta$-dependence of the Gilbert damping, leading to a switching behavior. This switching occurs at the critical voltage $V_{c,1}$, which can be determined by solving $\dot\phi\tilde g(\theta)+I_s(\theta,\dot\phi)=0$ for $\dot\phi=H_0$. The critical voltage at which the magnetization switches is then 
	\begin{equation}
		\Vcone = -H_0 \frac{\Gamma_\Sigma^2-\Gamma_\Delta^2}{2\Gamma_r\Gamma_\Delta}.\label{eq:Vcone}
	\end{equation}
	Note that we take $\Gamma_\Delta<0$ and thus $V_{c,1}/H_0>0$.
	\subsection{Presence of internal relaxation}
	In the presence of internal relaxation, i.e., for $\tau_{\bath}^{-1}\neq0$, the spin-transfer torque gains an additional contribution, $I_f(\theta,\dot\phi)$ [\cref{eq:results-sec:I4-intermediate}]. Additionally, the spin-dependent tunneling rates, $\Gamma_\sigma(\theta)$, which appear in $\Ib$ and $\Ih$ [\cref{eq:results-sec:Ib-intermediate,eq:results-sec:Ih-intermediate}] are also dependent on $\tau_{\bath}^{-1}$.
	
	\begin{figure*}
		\includegraphics{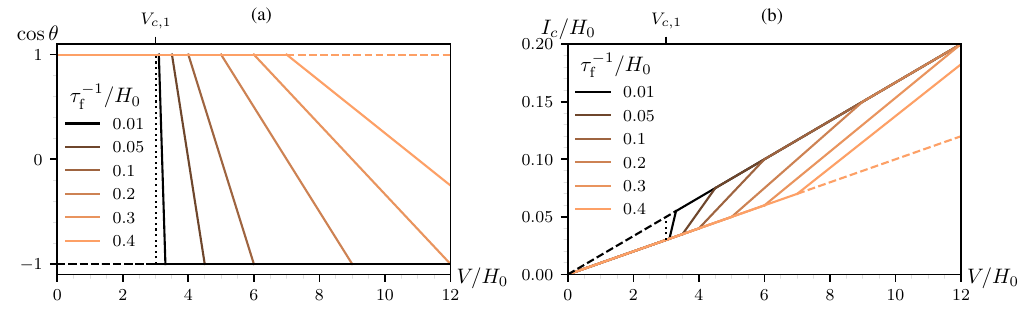}
		\caption{(a) The stable (solid) and unstable (dashed) steady-state solutions for increasing values of the internal relaxation rate $\tau^{-1}_{\bath}$. In absence of relaxation, $\tau_{\bath}^{-1}=0$, we obtain only a direct switching and no steady-state precessions, as indicated by the dotted vertical line. For finite relaxation, steady-state precessions are allowed. (b) The corresponding charge current through the nanomagnet. For reference, $\Gamma_\Sigma/H_0=0.2$.  \label{fig:costheta-current}}
	\end{figure*}

	We can now find solutions of $\dot\phi\tilde g(\theta) + I_s(\theta,\dot\phi)=0$ as
	\begin{equation}
		\cos\theta = \tau_{\bath} \frac{\Gamma_\Delta^2\rho_\Sigma - \Gamma_\Sigma\rho_\Delta(\Gamma_\Sigma + \tau_{\bath}^{-1})}{\Gamma_\Delta \rho_\Delta} - 2 \tau_{\bath}\frac{V}{H_0} \frac{\Gamma_r\rho_\Sigma}{\rho_\Delta} \label{eq:costheta-sol}
	\end{equation}
	and thus  steady-state solutions are now allowed. 
	These precessions start at the critical voltage
	\begin{equation}
		V_{c,1}^\bath = V_{c,1} -H_0\tau_{\bath}^{-1} \frac{\Gamma_\Sigma\rho_\Sigma + \Gamma_\Delta\rho_\Delta}{2\Gamma_r\Gamma_\Delta\rho_\Sigma} \label{eq:Vcrelaxation}
	\end{equation}
	and end at 
	\begin{equation}
		V_{c,2}^\bath =V_{c,1}^\bath + H_0 \tau_{\bath}^{-1}\frac{\rho_\Delta}{\Gamma_r\rho_\Sigma}
	\end{equation}
	such that for $V_{c,1}^\bath < V < V_{c,2}^\bath$ we have that $\cos\theta\neq\pm1$.
	
	We can thus draw the important conclusion that a finite internal relaxation, i.e. $\tau_f^{-1}\neq0$, allows for steady state precessions. This implies that internal relaxation, irregardless of its strength, is necessary in order to get an accurate description of the resulting macrospin dynamics---since in the absence of internal relaxation we only obtain switching and no steady state precessions. 
	
	Finally, the strength of the internal relaxation plays an important role in determining the voltage dependence of the $s_z$ component,
	\begin{equation}
		\frac{d(\cos\theta)}{d V} = - 2 \tau_{\bath} \frac{\Gamma_r\rho_\Sigma}{H_0\rho_\Delta}. \label{eq:slope}
	\end{equation}

	Furthermore, we find the charge current through the nanomagnet by inserting the steady-state solution [\cref{eq:costheta-sol}] in \cref{eq:Ic} to obtain
	\begin{equation}
		I_c=\tau_{\bath}H_0\frac{\Gamma_r\rho_\Sigma}{\Gamma_\Delta} \left[ 2\frac{V}{H_0}\Gamma_\Delta(\tau_{\bath}^{-1} + \Gamma_r) + (\tau_{\bath}^{-1}+\Gamma_\Sigma)^2-\Gamma_\Delta^2
		\right].
	\end{equation}

	We now show the resulting precessional states in \cref{fig:costheta-current}, for intermediate values of $\tau_f^{-1}$. The dashed lines indicate unstable states. First note that in the absence of relaxation, only switching is allowed at $V_{c,1}$, as indicated by the vertical dashed line. At a finite relaxation rate $\tau_{\bath}^{-1}$ we obverse the existence of precessions, with a slope which is proportional to $\tau_{\bath}$, as can be seen from \cref{eq:slope}. For a small relaxation rate $\tau_{\bath}^{-1}$ we therefore obtain precessions only over a small voltage range. 
	
	Increasing the relaxation rate further, the slope further increases, in addition to the critical voltage. This can be explained from the fact that the voltage bias needs to overcome the Gilbert damping enhancement, [\cref{eq:gtilde-final}], where the relaxation rate $\tau_{\bath}^{-1}$ also enters. We stress here that the relaxation rate $\tau_{\bath}^{-1}$ is typically not experimentally controllable, and thus an experiment that directly probes this change of slope or change in critical voltage as a function of $\tau_{\bath}^{-1}$ is not easily realized. 
	
	In the limit of strong relaxation, where $\tau_{\bath}^{-1}$ is larger than any other tunneling rate we obtain a very shallow slope, starting at a very high critical voltage. In fact, the limit of $\tau_{\bath}^{-1}\rightarrow\infty$ is ill-defined, requiring an infinite voltage to induce precessions of the macrospin direction. This corresponds to the fact that for an infinite relaxation rate, all spin will have relaxed to an equilibrium distribution on the nanomagnet before being transferred to the macrospin, and thus no dynamics are possible.

	In addition to the precessional states, we also show the resulting charge current flowing between the attached left and right lead through the nanomagnet in \cref{fig:costheta-current}. For voltages below the critical current, we obtain an Ohmic relation between the applied voltage and current, with two different slopes [and thus different resistances] corresponding to the stable solution on the north pole, and the unstable solution on the south pole.	Upon reaching the critical voltage, the stable solution increases its slope, thus effectively reducing the resistance of the nanomagnet over the voltage range where precessions are allowed.

	We conclude that internal relaxation plays an important role in the qualitative and quantitative description of steady-state precessions in a spin-torque oscillator. The precise value of the relaxation rate $\tau_{\bath}^{-1}$ can be estimated by noting that the Gilbert damping in absence of attached leads is in our formalism proportional to $\rho_\Sigma\tau_{\bath}^{-1}$. Once leads are attached, the Gilbert damping gains a contribution of $\rho_\Sigma\Gamma_\Sigma-\rho_\Delta\Gamma_\Delta\cos\theta$, where the $\cos\theta$-dependency allows experimental measurements to determine $\rho_\Sigma\Gamma_\Sigma$ and $\rho_\Delta\Gamma_\Delta$ separately. From these measurements of the Gilbert damping enhancement we know that $\rho_\Sigma\Gamma_\Sigma$ is roughly comparable to the ``bare'' Gilbert damping \cite{tserkovnyakEnhancedGilbertDamping2002}. We can thus conclude that $\Gamma_\Sigma/\tau_{\bath}^{-1}\approx1$, although we stress that this is a very rough estimate, which is strongly dependent on system parameters. For comparison, in \cref{fig:costheta-current} this corresponds to $\tau_{\bath}^{-1}/H_0=0.2$.

	\section{Steady state solutions in presence of anisotropy}
	\label{sec:anisotropy}
	Of particular interest is the behavior of the system in the presence of magnetic anisotropy, which can be used to model a wide range of spin-torque oscillators. We have included an uni-axial magnetic anisotropy and choose $K/H_0=0.2$ and also include a bulk Gilbert damping, $\alpha_0=10^{-3}$ for completeness. In the presence of anisotropy the deviation from the $z$-axis $\cos\theta$ is coupled to the precession frequency $\dot\phi$ and we therefore numerically solve the LLGS equation [\cref{eq:LLGS2}].
	We set the spin initially along the north pole, with a small deviation, and the LLGS equation is numerically solved for a period of $t_N\times H_0=10^4$, after which the position of the spin along the $z$-axis is recorded. We therefore only get access to the stable precessions, and do not gain information about the unstable precessions.

	\begin{figure}
		\includegraphics[width=\columnwidth]{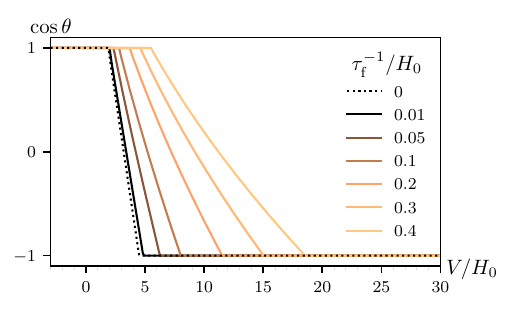};
		\caption{In the presence of anisotropy, the numerically obtained stationary solutions for varying values of the spin relaxation rate $\tau_{\bath}^{-1}$. The dotted black line is the $\tau_\bath^{-1}=0$ solution. For reference, $\Gamma_\Sigma/H_0=0.2$.  \label{fig:intermediate-anistropy-up}}
	\end{figure}

	We obtain the deviation from the $z$-axis, $\cos\theta$ as a function of applied bias voltage $V$ and show $\cos\theta$ in Fig.~\ref{fig:intermediate-anistropy-up}. 
	Comparing the results with the zero anisotropy case in \cref{fig:costheta-current}, we observe an important qualitative difference for both kinds of relaxation: in the absence of internal relaxation, $\tau_{\bath}^{-1}=0$, precessions are now allowed. These precessions however follow a linear dependency on voltage, which is not in agreement with the generally accepted experimental evidence that oscillation power is a nonlinear function of applied voltage in the presence of anisotropy \cite{kiselevMicrowaveOscillationsNanomagnet2003,tiberkevichNonlinearPhenomenologicalModel2007}.

	For a finite relaxation rate $\tau_\bath^{-1}$, we however obtain a nonlinear relation between the applied voltage $V$ and $\cos\theta$. We note here that the degree of nonlinearity has been the subject of some discussion, and typically requires additional modifications to the conventional LLGS description \cite{tiberkevichNonlinearPhenomenologicalModel2007}. The introduction of the relaxation rate provides one possible way that this discrepancy can be addressed.

	We therefore conclude that in the presence of anisotropy, the relaxation rate plays an important role, both qualitatively by changing the linear to nonlinear onset of preccessions, and quantitatively by changing the degree of nonlinearity.

	\section{Relaxation introduced in the rotating frame}
	\label{sec:rotating}
	In the previous sections we have modeled spin relaxation processes by introducing a fictitious lead in the laboratory frame. This is consistent with the expectation that this relaxation is due to spin-orbit coupling, which couples the spin degree of freedom to the lattice, allowing for the dissipation of angular momentum within the laboratory frame \cite{aleinerSpinOrbitCouplingEffects2001}.
	
	In this section, we propose a secondary, \emph{ad-hoc}, approach to introduce spin relaxation for relaxation processes within the nanomagnet’s electron system. One could intuitively expect a system to relax to equilibrium in that frame of reference, where the Hamiltonian is (at least approximately) time independent. One would thus also reasonably expect that the electrons would relax to an equilibrium distribution within the rotating frame. In this section, we work out this possibility by introducing a fictitious lead within the rotating frame, where the effective Hamiltonian is (approximately) constant. We stress here that we introduce this relaxation \emph{ad hoc}, i.e., without microscopic justification. In what follows, we will rederive the macrospin dynamics, essentially redoing (parts) of \cref{sec:method}. This can be skipped if needed, and we give the macrospin dynamics in \cref{sec:rotating-frame:macrospin-dynamics}.
	
	We obtain, after integrating out the fermionic fields and re-exponentiating, an action similar to \cref{eq:iS-rotating}:
	\begin{equation}
		i\mathcal{S}_{\Psi} = \Tr\ln\left[G_0^{-1} + JS\frac{\sigma^z}{2}-R^\dagger\Sigma R-\Sigma_{\rf}- Q\right].
	\end{equation}
	Here, to model relaxation processes in the rotating frame, we have introduced the self-energy $\Sigma_{\rf}$ that represents the interaction with a fictitious lead in the rotating frame. It is diagonal in the rotating frame, whereas the left and right self-energy is rotating through the rotation matrix $R$.
	The self-energy due to interaction with the fictitious lead is given by $\Sigma_{\rf}^{R/A}(\omega)=\mp \ii\tau_{\rf}^{-1}$ and $\Sigma_{\rf}^K(\omega)=-2\ii\tau_{\rf}^{-1} F_{\rf} (\omega)$, where $\tau_{\rf}$ is the relaxation rate resulting from the interaction with this fictitious lead and $F_{\rf}(\omega)=\tanh((\omega-\mu_{\rf}) / 2T_{\rf})$. Again, we will self-consistently determine $\mu_{\rf}$ by requiring that the charge current between the nanomagnet and the fictitious lead is zero. This charge current is independent of whether we introduce the bath in the rotating frame or the laboratory frame, since the charge current is rotationally invariant. The charge current is thus given by \cref{eq:Ic-bath} and $\mu_{\rf}$ is identical to $\mu_{\bath}$ [\cref{eq:mu3-nolim}], except, of course, for the replacement of $\tau_{\bath}^{-1}$ by $\tau_\mathrm{rf}^{-1}$.
	
	We then obtain that the slow distribution function [analogous to \cref{eq:slow-dist-function}] is given by
	\begin{equation}
		\begin{split}
			\tilde F_{s}^{\sigma}\left(\bar{t},\omega\right)=&\frac{1}{\Gamma_{\sigma}\left(\theta_{c}\right)} \Bigr[\cos^{2}\frac{\theta_{c}}{2}\Gamma_{l}^{\sigma}F_{l}\left(\omega+\sigma\omega_{-}\right) \\
			&+\sin^{2}\frac{\theta_{c}}{2}\Gamma_{l}^{\bar{\sigma}}F_{l}\left(\omega+\bar{\sigma}\omega_{+}\right)\\
			&+\cos^{2}\frac{\theta_{c}}{2}\Gamma_{r}F_{r}\left(\omega+\sigma\omega_{-}\right)\\
			&+\sin^{2}\frac{\theta_{c}}{2}\Gamma_{r}F_{r}\left(\omega+\bar{\sigma}\omega_{+}\right)\\
			&+\frac{1}{\tau_{\rf}} F_{\rf}\left(\omega\right)\Bigr].
		\end{split}
	\end{equation}
	
	Following the same steps as before, we obtain a kernel analogous to \cref{eq:I4sigmasigma} as
	\begin{equation}
		I_{\rf}^{\sigma\sigma'}  =\frac{1}{\tau_{\rf}}\frac{2\rho_{\sigma}}{\Gamma_{\sigma}\left(\theta_{c}\right)}\left(\Gamma_{l}^{\sigma'}\left(\mu_{l}-\mu_{\rf}\right)+\Gamma_{r}\left(\mu_{r}-\mu_{\rf}\right)\right) \label{eq:I3sigmasigma}
	\end{equation}
	and the resulting torques are given by summation over the spin indices [defined in \cref{eq:Isummation}] as
	
	\begin{align}
		\Ib[\rf]& =\frac{2\Gamma_{\Delta}\Gamma_{r}}{\Gamma_{\uparrow}^{\rf}\left(\theta\right)\Gamma_{\downarrow}^{\rf}\left(\theta\right)}\tilde{g}\left(\theta\right)V, \label{eq:rotating:Ib-intermediate}\\
		\Ih[\rf]& =\frac{\Gamma_{\Delta}^{2}}{\Gamma_{\uparrow}^{\rf}\left(\theta\right)\Gamma_{\downarrow}^{\rf}\left(\theta\right)}\tilde{g}\left(\theta\right)\frac{1}{2}\sin^2\theta\,\dot\phi, \label{eq:rotating:Ih-intermediate}\\
		I^{\rf}(\theta,\dot{\phi})&=\frac{\rho_\downarrow \Gamma^{\rf}_\uparrow(\theta) +\rho_\downarrow \Gamma^{\rf}_\downarrow(\theta) }{\tau_{\rf}\Gamma_{\downarrow}^{\rf}\left(\theta\right)\Gamma_{\uparrow}^{\rf}\left(\theta\right)} \Gamma_\Delta\left(\mu_l - \mu_{\rf}(\theta,\dot\phi) \right). \label{eq:I3-intermediate}
	\end{align}
	where $\Gamma^{\rf}_{\sigma}(\theta)\equiv\cos^2\frac{\theta}{2} \Gamma_l^\sigma + \sin^2\frac{\theta}{2} \Gamma_l^{\bar{\sigma }}+ \Gamma_r +\tau_{\rf}^{-1}$ are the spin-dependent level broadenings, which are importantly dependent on $\tau_{\rf}^{-1}$.
	
	Finally, the bath introduced within the rotating frame does not affect the Gilbert damping enhancement, and thus $\tilde g(\theta)=\frac12\left[\rho_\Sigma \Gamma_\Sigma - \rho_\Delta \Gamma_\Delta \cos\theta\right]$.
	
	\subsection{Macrospin dynamics}
	\label{sec:rotating-frame:macrospin-dynamics}
	We now consider the macrospin dynamics resulting from the fictitious lead introduced in the rotating frame. Firstly, in the limit of no relaxation [$\tau_{\rf}^{-1}=0$] the contribution from the bath vanishes, and $I_{\rf}(\theta,\dot\phi)=0$. We then recover the switching behavior as discussed in \cref{sec:no-internal-relaxation}. 
	
\begin{figure}
		\includegraphics[width=\columnwidth]{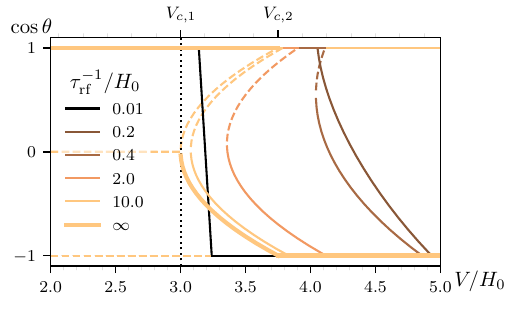}
			\caption{For a relaxation lead introducing in the rotating frame, the stable (solid) and unstable (dashed) precessions as a function of applied voltage . We show values of $\tau_{\rf}^{-1}$ ranging from no internal relaxation ($\tau_{\rf}^{-1}=0$, switching at $V_{c,1}$ indicated by the dotted line) to strong internal relaxation ($\tau_{\rf}^{-1}\rightarrow\infty$, thick line). For reference, $\Gamma_\Sigma/H_0=0.2$. 
			\label{fig:intermediate-rotating-frame}}
	\end{figure}

	If $\tau_{\rf}^{-1}\neq0$, we have that $I_s(\theta,\dot\phi)=\Ib[\rf]+\Ih[\rf]+I_{\rf}(\theta,\dot{\phi})$, given by \cref{eq:rotating:Ib-intermediate,eq:rotating:Ih-intermediate,eq:I3-intermediate}. We now find solutions of $\dot\phi\tilde g(\theta)+I_s(\theta,\dot\phi)=0$ for $\dot\phi=H_0$, which we show in \cref{fig:intermediate-rotating-frame}, i.e., the equivalent of \cref{fig:costheta-current}.

	We now observe that increasing the relaxation rate  $\tau_{\rf}^{-1}$ first allows for precessions, in line with our previous results. Increasing the relaxation rate beyond a certain critical value however renders some precessional states on the northern hemisphere unstable (compare for example $\tau_{\rf}^{-1}/H_0=0.1$ and $0.2$). This is a marked difference between the lead introduced in the rotating frame and in the laboratory frame, where increasing the relaxation rate increased the slope of the precession as a function of voltage. 
	Here, we obtain that increasing the relaxation rate instead gives rise to precessions on the southern hemisphere which are stable. 
	
	In fact, if we assume the relaxation rate to be  stronger than any tunneling rate, such that  $\tau_{\rf}\Gamma_\eta^\sigma\ll1$ for any $\eta\in\{l,r\}$ and $\sigma\in\{\uparrow,\downarrow\}$, we find that $\Ib[\rf]=0$ and $\Ih[\rf]=0$, and the only remaining torque comes from 
	\begin{equation}
		I^{\rf}(\theta,\dot{\phi})=\frac{\rho_\Sigma^2 \Gamma_\Delta \Gamma_r}{2g_e(\theta)}V  -\frac{\rho_\Sigma^2 \Gamma_\Delta^2}{2g_e(\theta)} \frac{1}{2}\sin^2\theta \, \dot{\phi} + O(\tau_{\rf}\Gamma_\eta^\sigma) . \label{eq:Irf-strong-relaxation}
	\end{equation}
	Importantly, the torque is now fully determined by the interaction with the ficticious lead. Physically, this can be interpreted from the fact that if $\tau_{\rf}\Gamma_\eta^\sigma\ll1$, the electrons on the nanomagnet will interact with the fictitious lead on timescales much faster than the tunneling rates from and to the left and right lead. In this limit the fictitious lead therefore dominates the dynamics. Interestingly, we still obtain a Slonczewski-like torque, proportional to the applied bias, \emph{and} a self-induced torque, proportional to $\dot\phi$ (the first and second term in \cref{eq:Irf-strong-relaxation} respectively).
	
	We also show in \cref{fig:intermediate-rotating-frame} the stationary solutions obtained in this limit of $\tau_{\rf}\Gamma_\eta^\sigma\ll1$ (thick line). In this limit, no steady state precessions on the northern hemisphere are stable, and at the critical voltage $V_{c,1}$ [\cref{eq:Vcone}] steady-state precessions are allowed, but these are only stable on the southern hemisphere. The stable solutions follow a square-root dependency on the applied voltage, and end at a second critical voltage 
	\begin{equation}
		\Vctwo = -H_0\frac{ \Gamma^2_\Sigma \rho^2_\Sigma - \Gamma_\Delta^2\rho_\Delta^2}{2\Gamma_r \Gamma_\Delta \rho_\Sigma^2} \label{eq:Vctwo}.
	\end{equation}

	We want to highlight here that the resulting torque for strong relaxation, \cref{eq:Irf-strong-relaxation}, is intimately connected to our previous work \cite{gunninkChargeConservationSpintorque2024}, where we imposed charge conservation in combination with strong internal relaxation of the electrons in the nanomagnet---in a mostly classical formalism. The results obtained here in the limit of strong relaxation reduce to those obtained in Ref.~\cite{gunninkChargeConservationSpintorque2024}, by identifying the spin-conductance as $g_l^s=\rho_\Sigma \Gamma_\Delta$ \cite{sunMagnetoresistanceSpintransferTorque2008}, cf. compare the self-induced torque in \cref{eq:Irf-strong-relaxation} and Eq.~(10) in Ref.~\cite{gunninkChargeConservationSpintorque2024}. This thus serves as an important connection between these two results, even though a completely different formalism was used to obtain them.
	
	Going one step further and, on top of strong internal relaxation $\tau_\mathrm{rf} \Gamma_\eta^\sigma \ll 1$, assuming the right tunnel junction to be much more transparent than the left tunnel junction $\Gamma_r \gg \Gamma_l^\sigma$, we can reproduce (the noiseless part of) the spin dynamics in Ref.~\cite{chudnovskiySpinTorqueShotNoise2008}. Thus, our approach suggests that Ref.~\cite{chudnovskiySpinTorqueShotNoise2008} implicitly assumes strong internal relaxation. This explains why we could reproduce the spin dynamics of Ref.~\cite{chudnovskiySpinTorqueShotNoise2008} in Ref.~\cite{gunninkChargeConservationSpintorque2024}, where we assumed strong internal relaxation, but not in Ref.~\cite{ludwigStrongNonequilibriumEffects2017}, where we did not account for any internal relaxation mechanisms.
	
	 However, note that we need to introduce the fictitious lead \textit{ad hoc} in the rotating frame to reproduce the results of Refs.~\cite{gunninkChargeConservationSpintorque2024} and \cite{chudnovskiySpinTorqueShotNoise2008}. If we take the same limit of $\tau_{\rf}\Gamma_\eta^\sigma\ll1$ for the fictitious lead introduced in the laboratory frame, we obtain that the Gilbert damping enhancement becomes infinite---and thus no steady-state precessions are possible.

	\section{Conclusion and discussion}
	\label{sec:conclusion}
	In conclusion, we have studied the nonequilibrium physics of spin-torque systems, developing a formalism to include spin relaxation. The inherent nonequilibrium character of a driven macrospin introduces a self-induced torque, which has mostly been overlooked; however, see \cite{ludwigStrongNonequilibriumEffects2017,gunninkChargeConservationSpintorque2024}. We have shown that in order to obtain steady-state precessions in the absence of magnetic anisotropy, a finite amount of spin relaxation needs to be considered. We also show how our formalism can be connected to previous results considering no spin relaxation \cite{ludwigStrongNonequilibriumEffects2017} and strong spin relaxation \cite{gunninkChargeConservationSpintorque2024}. Importantly, the self-induced torque is relevant irregardless of the strength of internal spin relaxation, highlighting that it should be considered in any real system.

	For the specific geometry we consider here, where the fixed magnetization is parallel to the free magnetization, we note that the vectorial dependency of the self-induced torque can also be written as 
	\begin{equation}
		\frac{1}{2}\sin^2\theta\,\dot\phi = \bm m_{\mathrm{fix}}\cdot(\bm s \times \dot{\bm{s}}).\label{eq:stt-vector}
	\end{equation}
	Therefore, in order to measure the self-induced torque experimentally, one can take advantage of this scaling, since the Slonczwewski torque has a different dependency on the relative orientation of the fixed and free magnetization. Thus, by varying the external magnetic field such that the precession moves from parallel to perpendicular to the magnetization of the lead, the self-induced current could be differentiated from the Slonczewski-like torque; compare Ref.~\cite{gunninkChargeConservationSpintorque2024}.

	The self-induced torque, as considered here, is fundamentally a result from the interplay between the magnetization dynamics and the left and right leads, combined with spin relaxation. This includes a wide class of spin-transfer torque devices, but does not include the spin-Hall nano-oscillators \cite{demidovMagneticNanooscillatorDriven2012, liuMagneticOscillationsDriven2012,chenSpinTorqueSpinHallNanoOscillators2016}, where the torque on the nanomagnet is  generated by a voltage in an adjacent heavy metal through the spin Hall effect. There is thus no voltage bias applied over the nanomagnet, and the setup as sketched in Fig.~\ref{fig:system} is no longer valid. Further work is therefore needed to understand if a self-induced torque is also present in these devices.

	In this work we have considered a zero-dimensional nanomagnet, but an extension to a one-dimensional magnet connected to leads would be an interesting development of our formalism. In particular, previous results have suggested that so-called geometric torques play an important role in driving magnetic textures, such as domain walls \cite{bajpaiSpintronicsMeetsNonadiabatic2020,osorioWideFerromagneticDomain2021,petrovicSpinChargePumping2018}. Similarily, the self-induced torque obtained in this work could play an important role in driving magnetic textures.
	
	\appendix
	
	\section{The Wess-Zumino-Novikov-Witten and Landau-Zener type action}
	\label{app:deltaQ}
	When expanding the action in quantum components, we obtained a contribution to the action proportional to $\delta\tilde Q$ in \cref{eq:S_Q}. This can be further split by writing $Q=Q_\parallel + Q_\perp$, with 
	\begin{align}
		Q_\parallel&=\left[\dot \phi (1-\cos \theta) - \dot \chi \right] \frac{\sigma_z}{2}, \\ Q_\perp&=e^{-\chi\sigma_z}\left[\dot \phi  \sin \theta \frac{\sigma_x}{2} -\dot \theta \frac{\sigma_y}{2} \right] e^{-\phi\sigma_z},
	\end{align}
	which are diagonal and off-diagonal in spin space respectively. 
	
	We can thus also write $\delta \tilde{Q}=\delta\tilde{Q}_\parallel + \delta\tilde{Q}_\perp$ to obtain two contributions to the action
	\begin{align}
		\ii\mathcal{S}_{\mathrm{WZNW}} &= -\Tr\left[\tilde{G}_c \delta\tilde{Q}_\parallel \right],	\\
		\ii\mathcal{S}_{\mathrm{LZ}} &= -\Tr\left[\tilde{G}_c \delta\tilde{Q}_\perp \right], 
	\end{align}
	which are a Wess-Zumino-Novikov-Witten (WZNW) and Landau-Zener (LZ) type action.  
	
	The Wess-Zumino-Novikov-Witten (WZNW) type action accounts for the Berry phase of the delocalized (conduction band) electrons \cite{shnirmanGeometricQuantumNoise2015}. However, assuming the macrospin to consist mostly of the spins of localized electrons, the WZNW action will only yield a small renormalization of the macrospin length and thus it can be disregarded.\footnote{The spin-length renormalization can be determined as in Refs.~\cite{shnirmanGeometricQuantumNoise2015,ludwigStrongNonequilibriumEffects2017,ludwigThermallyDrivenSpin2019}.}
	
	The Landau-Zener type action describes transitions between up and down spins \cite{shnirmanGeometricQuantumNoise2015}. We assume the macrospin in the nanomagnet to be the largest energy scale, which suppresses the off-diagonal contributions and we can thus disregard $\mathcal{S}_{\mathrm{LZ}}$. Physically speaking, this corresponds to the assumption that the spin of the electrons follows the macrospin adiabatically. Moreover, we use the gauge freedom to eliminate the classical part of $\tilde{Q}_{\parallel}$, by choosing $\dot{\chi}_c=\dot{\phi}_c(1-\cos\theta_c)$ and ${\chi}_q={\phi}_q(1-\cos\theta_c)$ \cite{shnirmanGeometricQuantumNoise2015,ludwigThermallyDrivenSpin2019}. 
	\section{Stability}
	\label{app:stability}
	In order to determine the stability of the solutions $\theta_0$, we expand as $\theta_0+\delta\theta$. Since $\partial_t \theta_0=0$ by construction, we have
	\begin{equation}
		\partial_t \delta\theta = a \delta\theta + b \partial_t{\delta\theta} + O(\delta\theta, \partial_t \delta\theta),
	\end{equation}
	where we have expanded up to first order in $\delta\theta, \partial_t \delta\theta$. The coefficients $a, b$ are complex functions of $\theta_0$, which we determine by expanding Eq.~\eqref{eq:LLGtheta} up to first order.
	The stability requirement can then be derived by noting that if $a / (1-b) <0$, fluctuations are damped out and thus the corresponding $\theta_0$ is stable.

	\section{List of definitions}
	\label{app:definitions}
		For completeness, we reproduce here a list of defined symbols:
		\begin{equation}
\Gamma_l^\sigma(\theta) \equiv \cos^2\frac{\theta}{2} \Gamma_l^\sigma + \sin^2\frac{\theta}{2}\Gamma_l^{\bar\sigma}
		\end{equation}
		is the level broadening of the left lead in the rotating frame.
	\begin{equation}
		\Gamma_\sigma\left(\theta\right)\equiv\Gamma_l^\sigma(\theta) + \Gamma_r +\tau_{\bath}^{-1}
	\end{equation}
	is the level broadening in the rotating frame.
	\begin{equation}
		\tilde{g}\left(\theta\right)\equiv\frac{1}{2}\left[\rho_{\Sigma}(\Gamma_{\Sigma}+\tau_{\bath}^{-1})-\rho_{\Delta}\Gamma_{\Delta}\cos\theta\right]
	\end{equation}
	is the spin conductance.      
	\begin{equation}
		g_e\left(\theta\right)\equiv\frac{1}{2}\left[\rho_{\Sigma}\Gamma_{\Sigma}+\rho_{\Delta}\Gamma_{\Delta}\cos\theta\right]
	\end{equation}
	is the electric conductance for electrons leaving the nanomagnet into the leads---it should not be confused with the electric conductance for electrons flowing through the nanomagnet from one lead to the other.

	Finally, we have defined 
	\begin{equation}
		\Gamma_\Delta \equiv (\Gamma_l^\uparrow - \Gamma_l^\uparrow) / 2;\quad \Gamma_\Sigma \equiv (\Gamma_l^\uparrow - \Gamma_l^\uparrow) / 2 + \Gamma_r
	\end{equation}
	\begin{equation}
		\rho_\Sigma \equiv \rho_\uparrow + \rho_\downarrow;\quad \rho_\Delta \equiv \rho_\uparrow - \rho_\downarrow
	\end{equation}
	Here $\tau_{\bath}^{-1}$ is the relaxation rate, $\Gamma_l^\sigma$ is the tunneling rate to the left lead for spin $\sigma$ and $\Gamma_r$ is the tunneling rate to the right lead, which is spin-independent. Finally, $\rho_{\sigma}$ is the density of states on the nanomagnet for spin $\sigma$.

		\begin{acknowledgments}
			We would like to thank Igor S. Burmistrov for fruitful discussions at the initial stages of the project. R.~D. is member of the D-ITP consortium, a program of the Dutch Research Council (NWO),  funded by the Dutch Ministry of Education, Culture and Science (OCW). This work is in part funded by the Fluid Spintronics research program with Project No.~182.069, financed by the Dutch Research Council (NWO), and by the Deutsche Forschungsgemeinschaft (DFG, German Research Foundation) -- Project No.~504261060 (Emmy Noether Programme). A.~S. acknowledges funding from the DFG Project SH 81/7-1. P.~G. acknowledges financial support from the Alexander von Humboldt postdoctoral fellowship.
		\end{acknowledgments}
		The data that support the findings of this article are openly available \cite{gunninkNonequilibriumEffectsSpintorque2025a}.

	\end{document}